\documentclass[conference]{IEEEtran}

\usepackage{graphicx}
\usepackage[export]{adjustbox}
\usepackage{tikz}
\usepackage{tabularx}
\usepackage{multirow}
\usepackage{siunitx}
\usepackage{enumitem}
\usepackage{amssymb}
\usepackage{tabularray}
\UseTblrLibrary{varwidth}
\usepackage{caption}
\usepackage{fancyhdr}
\ifCLASSOPTIONcompsoc
  \usepackage[nocompress]{cite}
\else
  \usepackage{cite}
\fi

\ifCLASSINFOpdf

\else

\fi

\usepackage[acronym]{glossaries}
\newacronym{Mem}{Mem}{Model Size}
\newacronym{PT}{PT}{Processing Time}

\newacronym{AODV}{AODV}{Adhoc On Demand Vector}
\newacronym{RFE}{RFE}{Recursive Feature Elimination}
\newacronym{MI}{MI}{Mutual Information}
\newacronym{SC}{SC}{Smart Contract}
\newacronym{GA-SVM}{GA-SVM}{ Genetic Algorithm-based Support Vector Machine} 
\newacronym {GA-DT}{GA-DT}{Genetic Algorithm-based Decision Tree}
\newacronym {MLP}{MLP}{Multilayer Perceptron}
\newacronym{KNN}{KNN}{K-Nearest Neighbors}
\newacronym{IPFS}{IPFS}{Interplanetary File System}
\newacronym{CompNB}{CompNB}{Complement NB}
\newacronym{NC}{NC}{Nearest Centroid}
\newacronym{VBFT}{VBFT}{Verifiable Byzantine Fault Tolerance}
\newacronym{BFT}{BFT}{Byzantine Fault Tolerance}
\newacronym{OCE}{OCE}{Ontology Consensus Engine}
\newacronym{VRF}{VRF}{Verifiable Random Function}

\newacronym{TSS}{TSS}{Transmitted Signal Strength}
\newacronym{TS}{TS}{Trust Score}
\newacronym {MN}{MN}{monitor node}
\newacronym {MITM}{MITM}{Man-in-the-Middle}
\newacronym {RSS}{RSS}{Received Signal Strength}
\newacronym {PSR}{PSR}{Packet Sending Rate}
\newacronym {PFR}{PFR}{packet forwarding rate} 
\newacronym {FD}{FD}{forwarding Delay }
\newacronym {ECA}{ECA}{Energy Consumption Amount}
\newacronym  {NF}{NF}{Node Affinity}
\newacronym{GBCRP}{GBCRP}{GAN and Block Chain based
secured Routing Protocol} 
\newacronym{GBDT}{GBDT}{gradient boosting decision tree}

\newacronym{SOA}{SOA}{service-oriented architecture}
\newacronym{RNN}{RNN}{recurrent neural networks}
\newacronym{VVLC}{VVLC}{Vehicular Visible Light Communication}
\newacronym{CM}{CM}{cluster member}
\newacronym{V2X}{V2X}{vehicle to everything communications} 
\newacronym{V2V}{V2V}{vehicle to vehicle} 
\newacronym{LTE}{LTE}{Long-Term Evolution}
\newacronym{DSRC}{DSRC}{dedicated short-range communications}
\newacronym{OWC}{OWC}{optical wireless communication}
\newacronym{FSO}{FSO}{free spcae optical communication}
\newacronym{PPM}{$M$-PPM}{multilevel pulse position modulation}
\newacronym{GPS}{GPS}{Global Positioning System}
\newacronym{DC-OFDM}{DCO-OFDM}{direct-current-optical-\gls{OFDM}}
\newacronym{PAM}{$M$-PAM}{pulse amplitude modulation}
\newacronym{EPPM}{EPPM}{Expurgated pulse position modulation}
\newacronym{PAPR}{PAPR}{peak to average power ratio}
\newacronym{DPPM}{DPPM}{Differential pulse position modulation}
\newacronym{OPPM}{OPPM}{Overlapping pulse position modulation}
\newacronym{VPPM}{VPPM}{variable PPM}
\newacronym{ISI}{ISI}{inter-symbol interference}
\newacronym{MPPM}{MPPM}{Multiple pulse position modulation}
\newacronym{FDGAN}{FDGAN}{Fully Distributed Generative
Adversarial Networks}
\newacronym{GOSS}{GOSS}{Gradient-Based One-Side Sampling} 

\newacronym{EFB}{EFB}{Exclusive Feature Bundling}
\newacronym{VPFT}{VPFT}{Verifiable Byzantine Fault Tolerance} 
\newacronym{IFS}{IFS}{Interplanetary File System}
\newacronym{OSI}{OSI}{Open Systems Interconnection}
\newacronym{DoS}{DoS}{Denial of Service}
\newacronym{NFR}{NFR}{non-functional requirement}
\newacronym{MAC}{MAC}{Media Access Control}
\newacronym{IT}{IT}{Information Technology}
\newacronym{IoT}{IoT}{Internet of Things}
\newacronym{WSN}{WSN}{Wireless Sensor Network}
\newacronym{iid}{iid}{independent and identically distributed}
\newacronym{DDoS}{DDoS}{Distributed denial of service}
\newacronym{IDS}{IDS}{Intrusion Detection System}
\newacronym{BC}{BC}{Blockchain}
\newacronym{BS}{BS}{base station}
\newacronym{CH}{CH}{cluster head}
\newacronym{QoS}{QoS}{quality of service}
\newacronym{ML}{ML}{Machine Learning}
\newacronym{LR}{LR}{Logistic Regression}
\newacronym{NB}{NB}{Naive Bayes}
\newacronym{K-NN}{K-NN}{K-Nearest Neighbors}
\newacronym{SVM}{SVM}{Support Vector Machine}
\newacronym{DT}{DT}{Decision Trees}
\newacronym{ANN}{ANN}{Artificial Neural Network}
\newacronym{RF}{RF}{Random Forests}
\newacronym{DL}{DL}{Deep learning}
\newacronym{RL}{RL}{reinforcement learning}
\newacronym{DRL}{DRL}{deep reinforcement learning}
\newacronym{PoW}{PoW}{Proof-of-Work}
\newacronym{AN}{AN}{aggregating node}

\newacronym{P2P}{P2P}{peer-to-peer}

\newacronym{ECDSA}{ECDSA}{elliptic-curve digital signature algorithm} 
\newacronym{tsp}{tsp}{transactions per second}
\newacronym{CPU}{CPU}{central processing unit}

\newacronym{FL}{FL}{Federated Learning}
\newacronym{PPV}{PPV}{positive prediction value}
\newacronym{ERR}{ERR}{Error rate}

\newacronym{GM}{GM}{geometric Mean}
\newacronym{RMSE}{RMSE}{root mean square error}
\newacronym{NRMSE}{NRMSE}{normalized \gls{RMSE}}
\newacronym{ROC}{ROC}{receiver operating characteristics}
\newacronym{RAM}{RAM}{random access memory}
\newacronym{KB}{KB}{kilobytes}
\newacronym{BW}{BW}{bandwidth}
\newacronym{Acc}{Acc}{classification accuracy}
\newacronym{Pd}{$P_d$}{probability of detection}
\newacronym{Pfa}{$P_{fa}$}{probability of false alarm}

\newacronym{Pmd}{$P_{md}$}{probability of misdetection}
\newacronym{HECC}{HECC}{Hyperel-liptic Curve Cryptography}
\newacronym{SDN}{SDN}{software-defined  networking} 
\newacronym{LSTM}{LSTM}{long short-term memory}
\newacronym{PoA}{PoA}{ Proof of Authority}
\newacronym{PoS}{PoS}{Proof of Stake}
\newacronym{DPoS}{DPoS}{Delegated Proof of Stake}
\newacronym{PoC}{PoC}{Proof of Capacity}
\newacronym{PBFT}{PBFT}{practical Byzantine fault tolerance}

\newacronym{CA}{CA}{Certificate Authority}
\newacronym{AES}{AES}{Advanced Encryption Standard} \newacronym{GA}{GA}{Genetic Algorithm}

\newacronym{CNN}{CNN}{Convolutional Neural Network}
\newacronym{HMM}{HMM}{Hidden Markov Model}
\newacronym{DNN}{DNN}{Deep Neural Network}
\newacronym{GAN}{GAN}{Generative Adversarial Networks}
\newacronym{ADC}{ADC}{Analog to Digital Converter}
\newacronym{PoET}{PoET}{Proof of Elapsed Time}
\newacronym{tps}{tps}{transactions per second} 
\newacronym{CPS}{CPS}{Cyber-Physical Systems}
\newacronym{HGB}{HGB}{Histogram Gradient Boost}
\newacronym{LEACH}{LEACH}{Low Energy Adaptive Clustering Hierarchy protocol}

\usepackage{url}
\usepackage{tikz}
\usepackage{enumitem}
\usepackage{ellipsis}
\usetikzlibrary{calc}
\usetikzlibrary{decorations.pathreplacing,decorations.markings,shapes.geometric,shapes.symbols}
\definecolor{lava}{rgb}{0.81, 0.06, 0.13}
\definecolor{myblue}{HTML}{B0D7FF}

\usepackage[many]{tcolorbox}
\usepackage{makecell}
\usepackage{minted}
\usepackage{xcolor} 
\usepackage{amsmath}
\usepackage{esvect}
\usepackage{subfigure}
\usepackage[utf8]{inputenc}
\usepackage{fourier} 
\usepackage{array}
\usepackage{cleveref}
\usetikzlibrary{positioning}
\usetikzlibrary{shapes.arrows}
\usepackage{svg}
\usepackage{censor}
\tikzset{
    mybox/.style={rectangle,
        draw,
        rounded corners,
        minimum width=2cm,
        inner sep=5pt,
        align=center,
        minimum height=1cm
    },
    myarrow/.style={draw=black,
        fill=white,
        minimum width=1cm,
        single arrow
    },
    longarrow/.style={draw=none,
        shading=axis,
        left color=white,
        right color=gray!20!white,
        minimum width=1cm,
        single arrow,
        anchor=south
    }
    }
\usepackage[export]{adjustbox}
\usepackage{multirow}
\usepackage{listings}
 \graphicspath{Figures/}   
\begin{document}

\author{
\IEEEauthorblockN{
Shereen Ismail\IEEEauthorrefmark{1},
Eman Hammad\IEEEauthorrefmark{2},
William Hatcher\IEEEauthorrefmark{2},
Salah Dandan\IEEEauthorrefmark{3},
Ammar Alomari\IEEEauthorrefmark{1},
Michael Spratt\IEEEauthorrefmark{1}
}

\IEEEauthorblockA{\IEEEauthorrefmark{1}Merit Network, Inc., University of Michigan, Ann Arbor, MI 48108, USA}

\IEEEauthorblockA{\IEEEauthorrefmark{2}iSTAR Lab, Texas A\&M University, College Station, TX 77843, USA}

\IEEEauthorblockA{\IEEEauthorrefmark{3}School of Electrical Engineering and Computer Science, University of North Dakota, Grand Forks, ND 58202, USA}

}

\title{Merit Network Telescope: Processing and Initial Insights from Nearly 20 Years of Darknet Traffic for Cybersecurity Research}

\maketitle
\begin{abstract}
This paper presents an initial longitudinal analysis of unsolicited Internet traffic collected between 2005 and 2025 by one of the largest and most persistent network telescopes in the United States, operated by Merit Network. The dataset provides a unique view into global threat activity as observed through scanning and backscatter traffic, key indicators of large-scale probing behavior, data outages, and ongoing denial-of-service (DoS) campaigns. To process this extensive archive, coarse-to-fine methodology is adopted in which general insights are first extracted through a resource-efficient metadata sub-pipeline, followed by a more detailed packet header sub-pipeline for finer-grained analysis. The methodology establishes two sub-pipelines to enable scalable processing of nearly two decades of telescope data and supports multi-level exploration of traffic dynamics. Initial insights highlight long-term trends and recurring traffic spikes, some attributable to Internet-wide scanning events and others likely linked to DoS activities. We present general observations spanning 2006–2024, with a focused analysis of traffic characteristics during 2024.
\end{abstract}

\begin{IEEEkeywords}
Network telescope, Traffic metadata analysis, Packet header analysis, Darknet, Unsolicited network traffic, Internet background radiation, cybersecurity
\end{IEEEkeywords}

\IEEEpeerreviewmaketitle


\section{Introduction}
\label{sectionI}

Unsolicited Internet traffic, often referred to as Internet Background Radiation (IBR), consists of packets sent to unused or unassigned IP addresses. Such traffic offers a valuable lens into the behavior of global threat actors, distributed scanning operations, botnets, and denial-of-service campaigns \cite{10903791}. Network telescopes, also known as darknet sensors, are a foundational tool in capturing this traffic at scale.
This paper presents a longitudinal study of unsolicited traffic collected a nearly 20-year period (2005–2025) from one of the largest and longest-operating network telescope deployments in the United States, hosted at Merit Network.

Unlike many short-duration or narrowly scoped datasets, the Merit telescope archive spans over almost two decades of continuous packet captures, representing a unique resource for understanding the evolution of malicious Internet behavior. The network telescope was originally deployed over a \texttt{/8} IP block, providing broad visibility into scanning and backscatter traffic. In 2018, a resource optimization effort scaled it down to a \texttt{/13} block, now called ORION, consisting of 1856 \texttt{/24} subnets (around 500,000 dark IPs), representing a 60\% reduction in address space with important implications for trend analysis and traffic normalization.

A critical key to enable such processing for long-term insights requires a careful consideration of the amount of data and the ability of the methodology to zoom in and out temporally, and on the attribute level to facilitate investigating long-term research questions. Hence, full processing of all files, including packet headers and payloads would require extensive resources. Hence, in this work we follow a coarse-to-fine methodology where general insights are first extracted through a more resources efficient sub-pipeline that considers coarse attributes, followed by a second sub-pipeline that extracts finer details but is more resource extensive. Necessary temporal sampling strategies are also made possible in the second finer sub-pipeline, to balance insights and processing resources. Important to note here, that this approach can be further expanded to enable finer analysis by extracting additional packet headers.

Specifically, this study focuses on the extraction, indexing, analysis, and visualization of high-level metadata from compressed packet capture files  (\texttt{.pcap.gz}), using two complementary sub-pipelines within the ORION Network Telescope processing framework: 1) \textbf{high-level metadata sub-pipeline} leverages the \texttt{capinfos} utility to derive time-series attributes such as packet rates, throughput, file sizes, and data density. These metrics are ingested into a time-series database (InfluxDB) and then visualized via Grafana dashboards. 2) \textbf{packet header sub-pipeline}, employs \texttt{Apache Drill} to extract packet level attributes, including timestamps, source and destination IP addresses, port numbers, and TCP header flags. The extracted attributes are then fed into a relational database (MariaDB) and then visualized via Grafana. Together, the two sub-pipelines enable multi-level, dynamic visualization and exploration of network trends and anomalies over time.

    


A key focus of this work is characterization of Internet-wide scanning and backscatter activities, through the telescope collected data and extracted features. The study also expands on data integrity challenges inherent in long-term passive monitoring. The data processing enables the quantification  of corrupted or incomplete capture files, identification of outages due to operational or network-layer causes, and the documentation of temporal gaps or inconsistencies. Special attention is given to the impact of the \texttt{/8} to \texttt{/13} IP space transition, to better understand the transition (smaller dark IP addresses space for the telescope) impact and potential need for normalization. 

This work builds the foundation for more involved analysis to address challenges such as: 1) scalable and context-involved strategies for data retention (what to keep) and processing with optimal use of constrained resources (storage, computing), 2) optimized mechanisms for process-per-need via hierarchal approaches, and 3) identification of useful enhancements to data collection and measurements. This work aims to ultimately support future research in large-scale network measurement, cybersecurity observatories, and Internet threat intelligence.

The rest of this paper is organized as follows: \Cref{sectionII} reviews related work. \Cref{sectionIII} describes our methodology, detailing the two complementary sub-pipelines developed in this study for metadata extraction, time-series conversion, and dashboard visualization. \Cref{sectionIV} presents our key observations from the dataset at the high-level and packet-header and discuss . Finally, \Cref{sectionV} summarizes our key findings and outlines future directions for telescope-based Internet threat research.


















\section{Related Work}
\label{sectionII}


Similar large-scale network telescope platforms to Merit’s long-term dataset include the UCSD Network Telescope, the DarknetBR and NICTER-E projects, as well as distributed and any cast network telescopes. UCSD’s installation, for example, monitors a globally routed /9 and /10 IPv4 space, offering extensive coverage that makes it especially suitable for detecting wide spread events~\cite{caidatelescope, caidadatasets}. In comparison, smaller telescopes or those with a distributed layout can aggregate disparate address ranges for broader observational reach, but may face challenges such as clock synchronization and varying network characteristics across sites~\cite{irwinbaseline, mooretelescopes}. Platforms with large continuous IP blocks tend to observe a greater volume of background radiation and have faster detection times for global scanning events, while smaller/fragmented telescopes may miss brief or low-rate activity and require statistical compensation~\cite{mooretelescopes}. The duration and scale of data collection are critical factors: UCSD and Merit, for example, offer multi-year continuity, supporting robust trend and anomaly analysis—whereas short-term telescopes deliver only snapshots of Internet-wide phenomena~\cite{lessismore, mooretelescopes}.


Griffioen et al. \cite{griffioen2024have} studied the data from a large (roughly /16) network telescope over the period of 10 years . Their telescope collected data from various subnets equating to roughly a /16 block. Their setup captures packets using LibPCAP a popular and widely used library for both capturing and analyzing network traffic. IP events are determined and combined with GeoIP data and stored in two databases. Periodocily scrips are ran to calculate trends. Lee et al. \cite{lee2003detection} provide an empirical analysis of scanning behavior and find that 91\% of port scanners target IP addresses sequentially. Pang et al. \cite{pang2004characteristics} additionally find that port scanning is highly targeted to certain ports.

Durumeric et al. \cite{durumeric2014internet} show that the high-level metrics like the origin of scans remained constant, but also identify that there are large changes since previous studies such as drastic changes in targeted ports and a major surge in scanning traffic due to the advent of new tools that make Internet scanning more accessible. Ghiette et al. \cite{ghiette2016remote} identify a large bias in how well-known tools are used along with a large geographical bias in tool usage. Large biases also exist in scans targeting certain ports, with 77\% of scans to Microsoft Remote Desktop Protocol (RDP) originating from China in 2014. \cite{durumeric2014internet}

Richter and Berger \cite{richter2019scanning} show that only a small fraction of scans actively target the entire IPv4 space, but that these scans account for more than 27\% of all scanning traffic due to their size. Durumeric et al. \cite{durumeric2014internet} have also identified this imbalance, with 0.28\% of scans generating nearly 80\% of the traffic. They separate backscatter from scans by only selecting TCP frames with the SYN flag set \cite{izhikevich2022predicting}. Prior work in \cite{durumeric2014internet}, \cite{ghiette2016remote} demonstrated that different scanner software can be distinguished based on packet-level properties.

As a complementary security mechanism to network telescopes, honeypots actively engage adversaries by emulating vulnerable services. A prior micro-level study utilized a honeypot deployment at Merit Network to analyze malicious activity \cite{11103659}. That study examined botnet-driven login attempts and malware exploitation through detailed payload inspection and session analysis, yielding service-specific insights into attacker behavior. While honeypots provide valuable micro-level observations, our longitudinal telescope study offers a macro-level perspective on unsolicited Internet traffic, capturing scanning and backscatter events across a much larger address space and over a two-decade time span (2005–2025). Together, these approaches underscore the complementary roles of active and passive monitoring in understanding global threat activity.

\definecolor{dkgreen}{rgb}{0,0.6,0}
\definecolor{gray}{rgb}{0.5,0.5,0.5}
\definecolor{mauve}{rgb}{0.58,0,0.82}
\lstset{language=SQL,
  basicstyle={\small\ttfamily},
  belowskip=3mm,
  breakatwhitespace=true,
  breaklines=true,
  classoffset=0,
  columns=flexible,
  commentstyle=\color{dkgreen},
  framexleftmargin=0.25em,
  frameshape={}{yy}{}{}, 
  keywordstyle=\color{blue},
  numbers=none, 
  numberstyle=\tiny\color{gray},
  showstringspaces=false,
  stringstyle=\color{mauve},
  tabsize=3,
  xleftmargin =1em
}

\section{Methodology}
\label{sectionIII}
\begin{figure*}[h!]
  \centering
    \includegraphics[width=\linewidth]{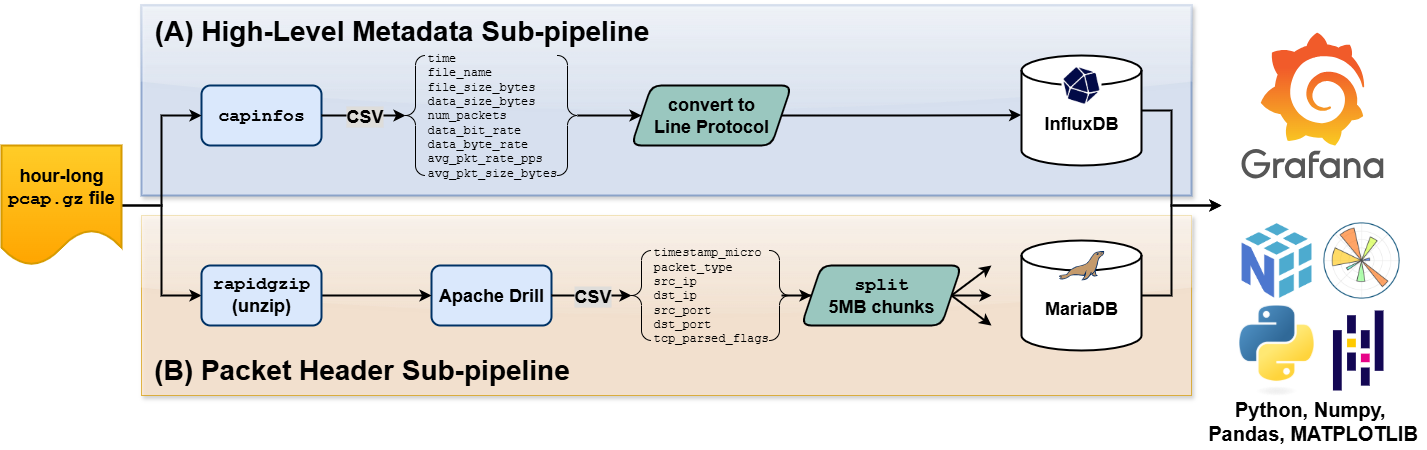}
    \caption{Coarse-to-Fine Sub-pipelines of the ORION Network Telescope}
    \label{fig:pipeline}
\end{figure*}


In this section, we expand on two aspects of this study: the coarse-to-fine sub-pipelines and database storage and visualization. The ORION network telescope workflow saves the captured internet traffic to a GNU-zipped (gzip) PCAP file every hour, with the file name in the format of \texttt{YYYY-MM-DD.HH.pcap.gz}. Each  compressed file size averages around 2 GB from 2006 to 2019 and 5 GB from 2020 to 2025. The industry standard library for processing pcap files "libpcap" from Wireshark is not multithreaded and hence would use as much RAM as possible until the process is killed by the operating system due to memory exhaustion. In this work, we developed two coarse-to-fine sub-pipelines that utilize libpcap alternatives for feature extractions that balance processing time and resource usage. 

\subsection{Coarse-to-Fine Sub-pipelines}
\label{sec:pipeline}

Two complementary sub-pipelines were developed for feature extraction from ORION network telescope traffic files: the first extracts high-level metadata from each .pcap.gz file; the second extracts the packet headers from unzipped .pcap files. \Cref{fig:pipeline} describes the developed and tested sub-pipelines. In what follows, Linux-based virtual machines were used and relevant commands are provided for reference.

Prior to the execution of the sub-pipelines, the telescope .pcap.gz files are copied from an archive storage to the server running the sub-pipelines. This vastly increases the performance of both sub-pipelines by reducing the access time of the files. We next expand on each sub-pipeline.

\subsubsection{High-Level Metadata Sub-pipeline} The high-level metadata sub-pipeline extracts summary statistics from raw \texttt{.pcap.gz} files to provide an overview of network traffic characteristics \Cref{fig:pipeline}(A). For each .pcap.gz file, it collects features, including:
\begin{itemize}
    \item time - timestamp of the latest packet captured. 
    \item file\_name - name of the .pcap.gz file.
    \item file\_size\_bytes - size of the .pcap.gz file in bytes.
    \item data\_size\_bytes - total amount of network traffic captured and stored.
    \item num\_packets - total number of packets captured.
    \item data\_bit\_rate and data\_byte\_rate - average speed of the network capture in bits per second and bytes per second, respectively
    \item avg\_pkt\_rate\_pps - average rate of packets per second
    \item avg\_pkt\_size\_bytes - average size of each packet in bytes
\end{itemize}

The high-level metadata sub-pipeline operates on each \texttt{.pcap.gz} file, which represents one hour of captured traffic, as follows:
\begin{enumerate}[label=\alph*), start=1]

\item \texttt{capinfos} is invoked with \texttt{capinfos -Tmr > out.csv} which saves the extracted metadata to a CSV file,
\item a bash script converts the CSV file to InfluxDB "Line Protocol", a text-based format for ingesting data into InfluxDB, and imports the data into InfluxDB.

\end{enumerate}
Features are stored in InfluxDB, a time-series database which can be quickly queried. The table occupies about 1GB on disk and uses 4GB of RAM. Completed processing for years 2006-2024 using this sub-pipeline demonstrates its ability to efficiently handle large volumes of traffic files, providing high-level insights into network activity and serving as a foundation for more detailed, specialized analyses.

\subsubsection{Packet Header Sub-pipeline} The packet header sub-pipeline extracts detailed packet-level information from pcap files for analysis (\Cref{fig:pipeline}(B)). Apache Drill is employed to extract packet headers from pcap files because of its robust performance with large files, ease of deployment, and ability to output CSV files. Apache Drill is an open source project which provides a SQL interface to many different types of data and data sources. It can be ran as a standalone server on a single machine or be deployed to multiple machines to serve the needs of many researchers and data scientists at once. It provides robust reporting capabilities as well~\cite{apachedrill}.

Apache Drill cannot, however, process compressed pcap files. Many other pcap processing tools (such as tshark) can natively process gzipped pcap files. Decompression is handled by \texttt{rapidgzip} as it is between 30 and 75 times faster than the standard Linux tool, \texttt{gzip} ~\cite{Knespel2025rapidgzip}. Extracted features are saved to MariaDB, an open source SQL database based on MySQL. MariaDB is not well suited to importing large files at once, so the CSV file is split into 5MB chunks using the Linux \texttt{split} command, then imported using the SQL query in \Cref{fig:mariadbImport}. Features extracted with Apache Drill include the following for each packet:
\begin{itemize}
    \item timestamp - timestamp of the packet with microsecond-level precision
    \item packet\_type - Layer 3 packet type: TCP, UDP, ICMP, or 'unknown')
    \item src\_ip and dst\_ip - source and destination IP addresses
    \item src\_port and dst\_port - source and destination port numbers
    \item tcp\_parsed\_flags - string of TCP flags separated by |
\end{itemize}

The packet header sub-pipeline can be described as follows, for each .pcap.gz file:
\begin{enumerate}[label=\alph*), start=1]
    \item \texttt{rapidgzip}\cite{Knespel2025rapidgzip} decompresses the .pcap.gz using multiple threads.
    \item Packet headers are extracted from Apache Drill using the SQL query in \Cref{fig:drillQuery}. Results are saved to a CSV file.
    \item The CSV file is split into 5MB chunks using the Linux \texttt{split} command
    \item The chunks are imported into MariaDB with 10 files imported in parallel with the \texttt{parallel} \cite{parallel} tool at once using the SQL query in \Cref{fig:mariadbImport}.
\end{enumerate}

\begin{figure}
  
    \begin{lstlisting}[language=SQL]
SELECT 
    timestamp_micro, 
    type AS packet_type, 
    src_ip, 
    dst_ip, 
    src_port, 
    dst_port, 
    tcp_parsed_flags 
FROM 
    dfs.`<PCAP-FILE>`;
\end{lstlisting}
\caption{Apache Drill SQL Query}
\label{fig:drillQuery}
\end{figure}

\begin{figure}
\begin{lstlisting}[language=SQL]
LOAD DATA LOCAL INFILE '<CSV-FILE>' 
INTO TABLE pcap_drill 
FIELDS TERMINATED BY ',' ENCLOSED BY '\'' 
IGNORE 1 LINES 
(@micro_epoch, packet_type, src_ip, dst_ip, src_port, dst_port, tcp_parsed_flags) 
SET timestamp_micro = FROM_UNIXTIME(@micro_epoch / 1000000);
\end{lstlisting}
\caption{MariaDB CSV Import SQL Query}
\label{fig:mariadbImport}
\end{figure}

The database houses over 7.74 billion packet header rows occupying 855GB of disk and uses less than 1GB RAM. The sub-pipeline can be customized to zoom in and to sample data from the ORION'S captured files in the period 2006-2024 as needed. For the purpose of this study, and as an illustrative example, the sub-pipeline was configured to extract the packet headers of all packets captured at the noon hour on Tuesdays for every week in the year 2024. 

\subsection{Database Storage \& Visualization}


For the packet-header sub-pipeline and the selected data scope (1 hour per week for 2024), over 7 billion rows are stored in MariaDB. To speed up querying, indexes are created on the \texttt{time}, \texttt{dst\_port}, and \texttt{src\_ip} columns. While database best practices recommend indexing frequently queried columns, they also caution against creating too many indexes, as this can degrade performance.





Grafana is configured to use both InfluxDB and MariaDB as data sources, providing a unified, queryable interface and dashboard builder. It supports near real-time visualization of high-level metadata stored in InfluxDB, while detailed packet header analytics from MariaDB are available in a batch-processing mode. Grafana enables interactive exploration, monitoring, and visualization across both sub-pipelines. User can create custom dashboards, apply dynamic filters, and execute complex queries to drill down into network traffic patterns, anomalies, and trends. In addition, Grafana’s alerting features allow to monitor key metrics and receive notifications for unusual activity, supporting both operational monitoring and exploratory research on ORION network telescope traffic data. In future work, dashboards could be shared across teams to enable collaborative analysis, which would further accelerate insights from large-scale network telemetry.

\begin{figure*} 
    \centering
    \includegraphics[width=\linewidth]{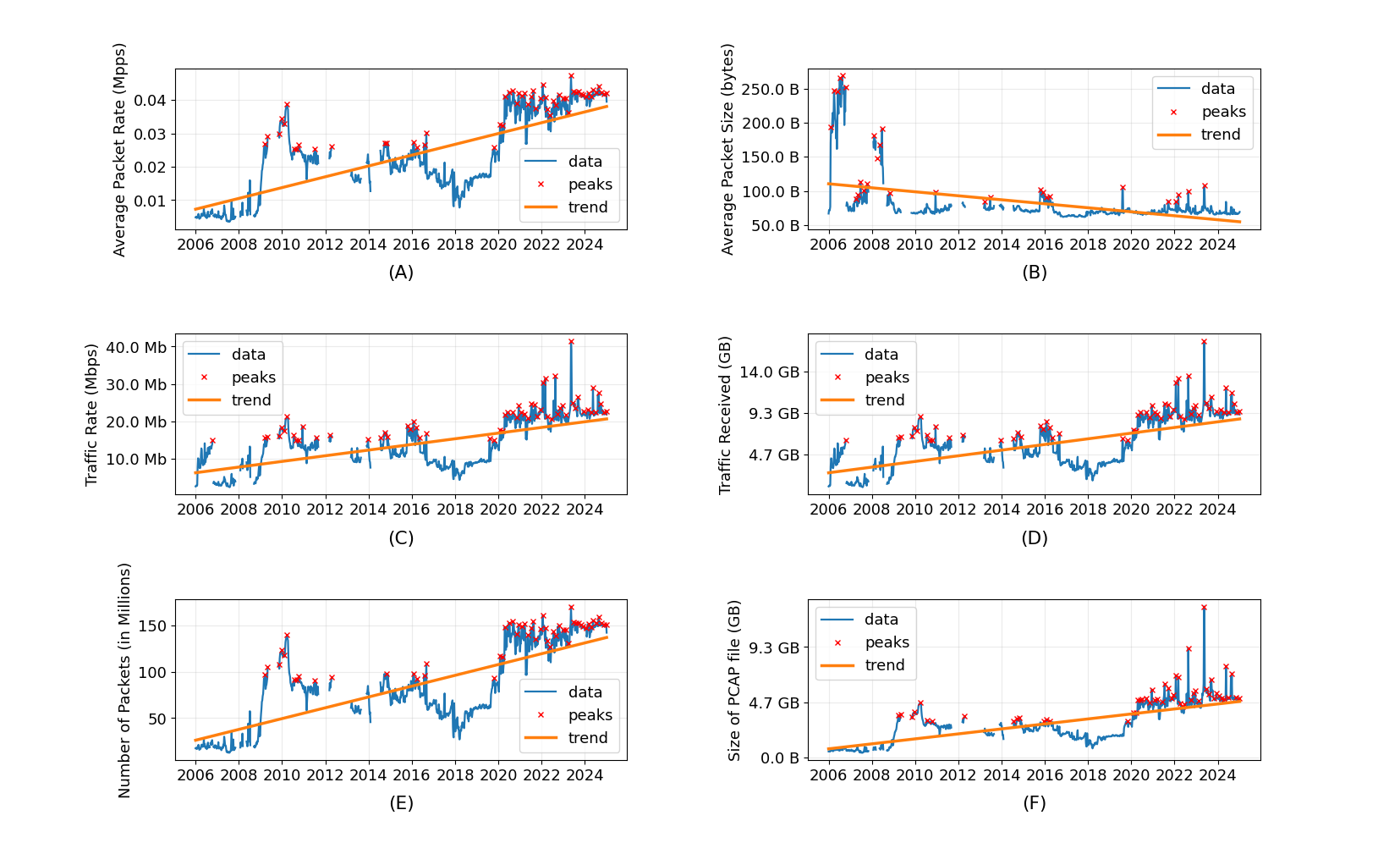}
    \caption{Longitudinal trends- and traffic spikes in ORION darknet data over the years 2006 to 2024.}
    \label{fig:15-yrs-trend}
\end{figure*}

\begin{figure*}
    \centering
    \includegraphics[width=0.9\linewidth]{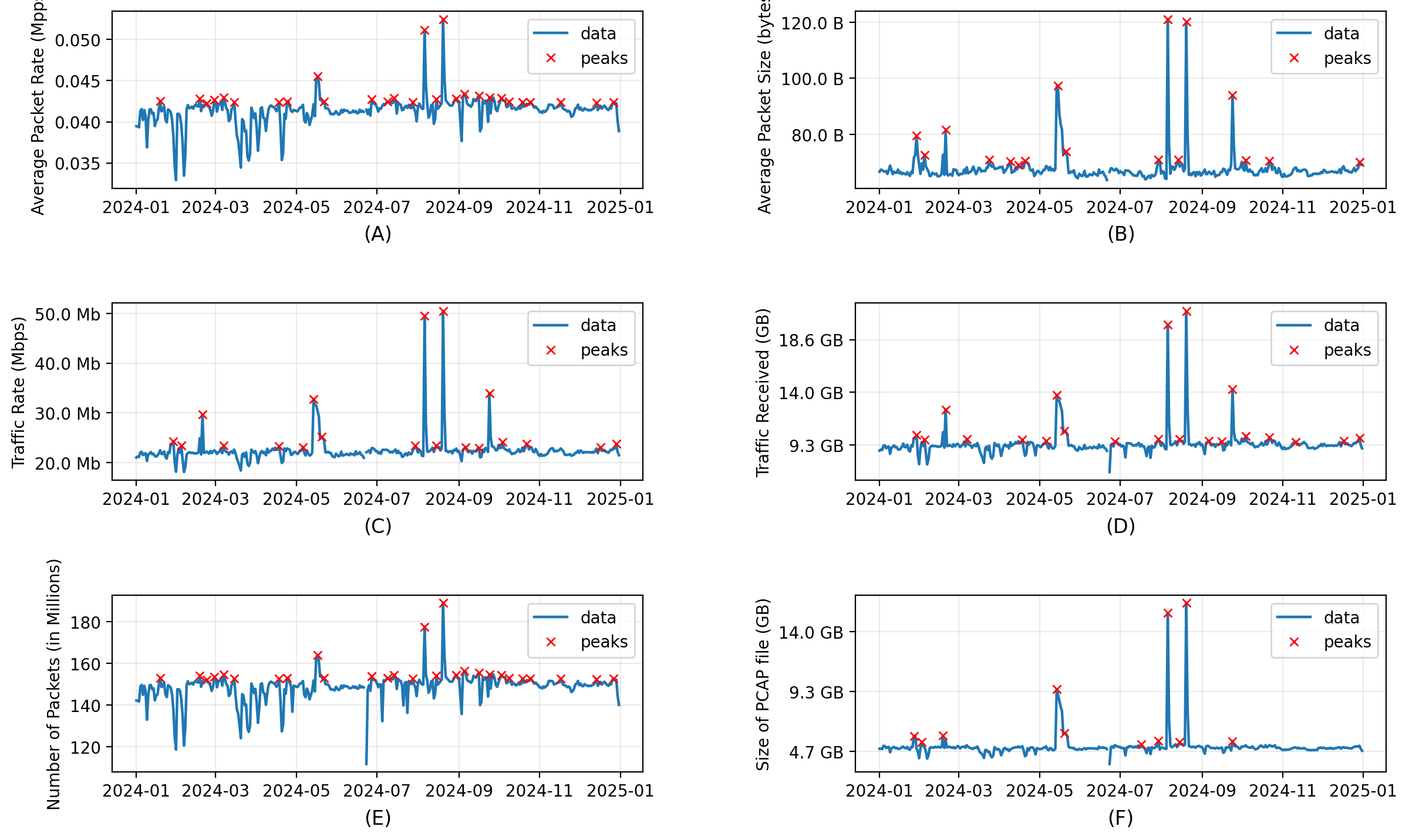}
    \caption{ Traffic trends and spikes in ORION darknet data in the year 2024.}
    \label{fig:2024-trends}
\end{figure*}

\section{Key Observations and Discussion}
\label{sectionIV}

\subsection{General High-Level Metadata Insights: }
The first pipeline allows us to observe trends and spikes in darknet traffic across the 2006-2024 years period as illustrated in \Cref{fig:15-yrs-trend}. \Cref{fig:2024-trends} zooms into the data gathered for the year 2024. Note that gaps in the chart represent data outages, which may occur either when the telescope was not running, when capture processes failed due to issues such as traffic spikes overwhelming the capture pipeline, or when files were corrupted or missing. The amount of traffic captured is observed to vary year-to-year, but remains more consistent within each year.


\Cref{fig:15-yrs-trend}(A) and \Cref{fig:2024-trends}(A) details the average packets per second captured by the network telescope in Mega packets per second; (B) depicts the average packet size in bytes; (C) visualizes the rate of traffic in Megabits per second (Mbps); (D) graphs the total amount of traffic captured within a month in GB; (E) shows how many packets captured in a month; (F) charts the size of each compressed PCAP file in GB. With the exception of (B) all graphs are closely correlated and have slight positive upward trends. While (B) has a negative linear trend, the average packet size is converging around 60--70 bytes. Further, investigation of actual packet payloads is needed to understand earlier trends of larger sizes during 2006-2008 versus the years after.
It is also noted that the average packet rate is growing faster than the traffic rate, hinting at more activity with relatively lower packet sizes which is more consistent with scanning and probing. 

\begin{figure}
    \centering
    \includegraphics[width=\columnwidth]{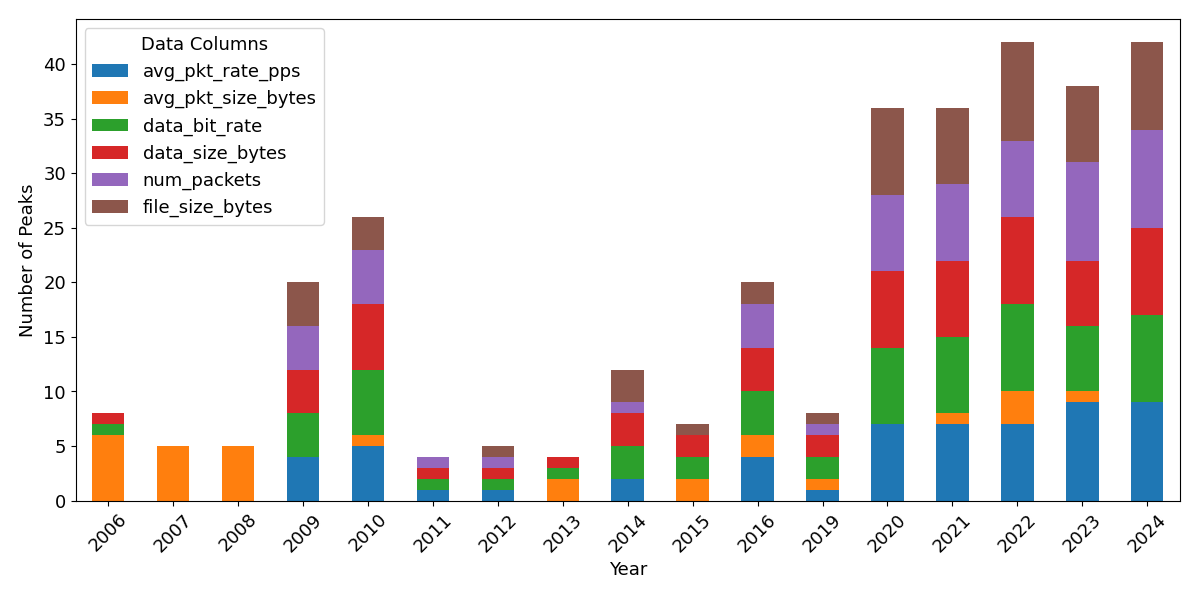}
    \caption{Number of peaks per year from 2006 to 2024}
    \label{fig:spikes-per-year}
\end{figure}

Peaks in the traffic data are found with SciPy's \texttt{find\_peaks} function with a height of 1.05 times mean of dataset and a minimum distance of 5 between peaks. \Cref{fig:spikes-per-year} shows the sum of peaks per year for each measurement. There exists a clear spike in 2009 and 2010, and a 2x increase starting in 2020. 

\vspace{0.3in}

\subsection{Packet-header Insights (year 2024):}

\begin{figure}
    \centering \includegraphics[width=\columnwidth]{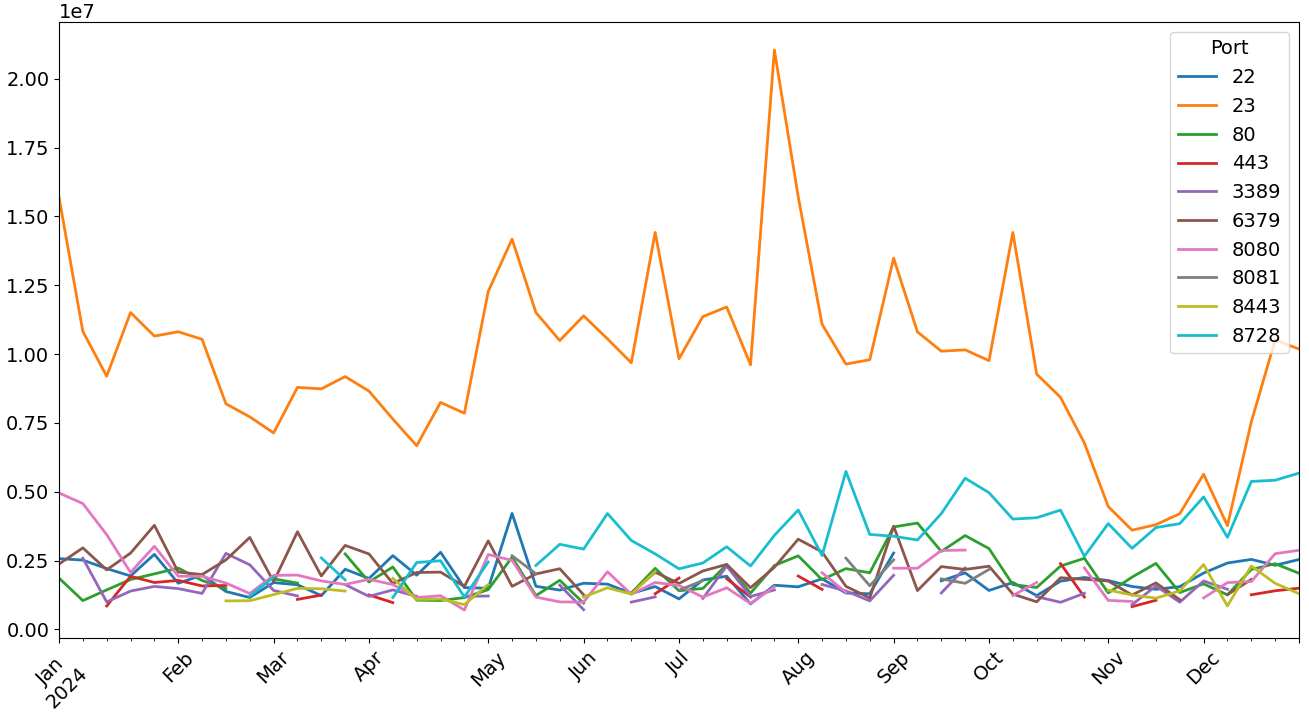}
    \caption{Top 10 Ports of the year 2024}
    \label{fig:top10ports_2024}
\end{figure}

Top 10 ports most frequently targeted in 2024 are shown in \Cref{fig:top10ports_2024}. Notably, port 23 (commonly used for Telnet) consistently overwhelmingly dominates the other ports by about a factor of 4, receiving over 20 million packets per month, which starkly outpaces the remaining ports and highlights its persistent attractiveness to attackers due to its historical vulnerabilities and prevalence on unsecured devices. The other ports, averaging about 3 million packets per month include MikroTik management (8728), Redis (6379), HTTP (80, 8080, 8081), HTTPS (443, and 8443), and SSH (22). The continued targeting of both legacy ports and those associated with modern web services reflects attackers' efforts focusing on exploiting outdated systems and probing popular cloud-facing applications. 

The Telnet port 23 dominance requires further study and comparisons with observations from similar telescopes or datasets.

\begin{figure}
    \centering
\includegraphics[width=\columnwidth]{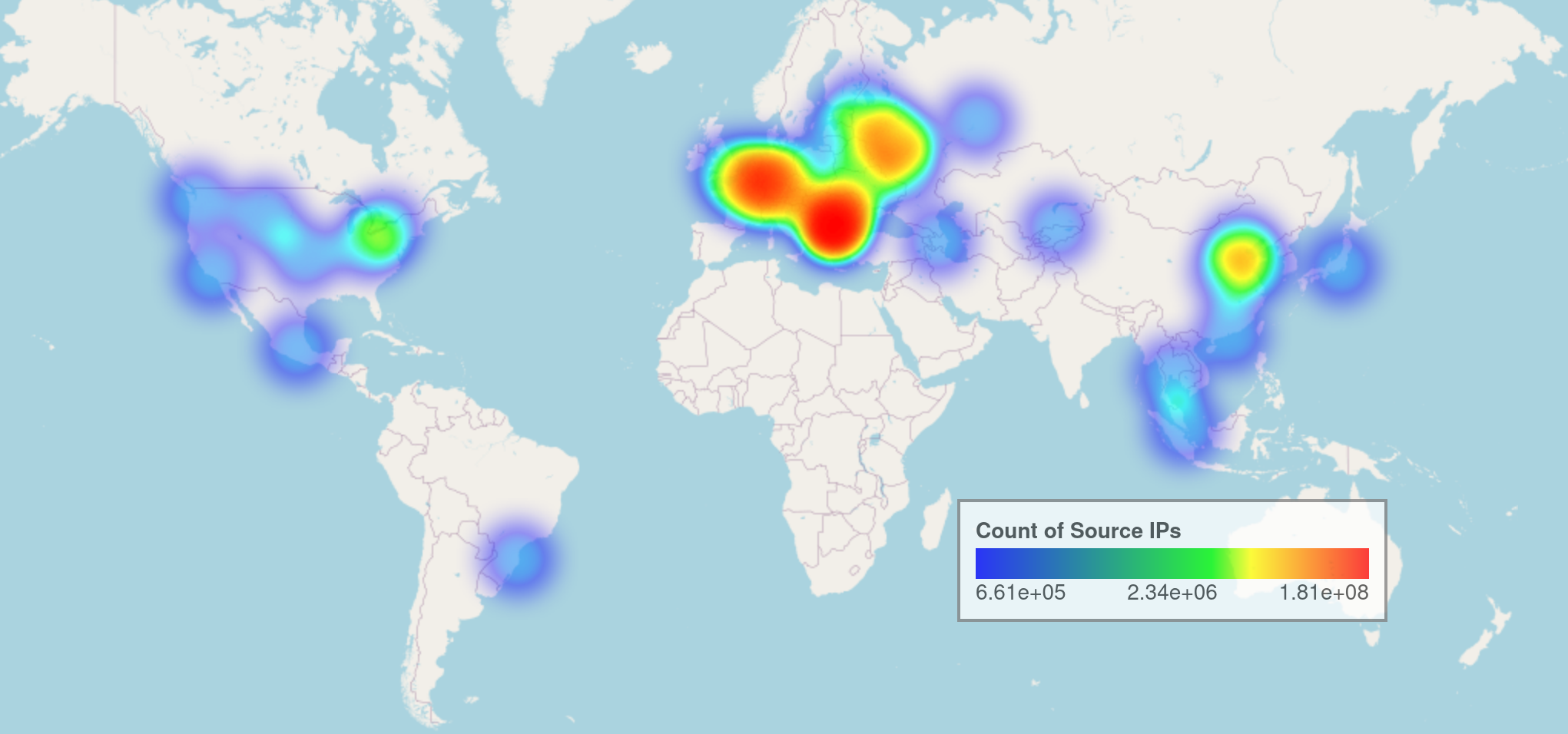}
    \caption{Top 10 Source IPs of the year 2024}
    \label{fig:top10ip_2024}
\end{figure}

The top 10 source IPs observed in 2024 are graphed in \Cref{fig:top10ip_2024} as a world heat-map. The most prominent hotspots are situated in Central and Eastern Europe which suggests that a substantial fraction of the top observed IPs are concentrated in this geographical corridor, potentially implicating these regions as focal points for either legitimate large-scale network operations or coordinated anomalous activities.

\subsection{Data Outages: }
\begin{figure}
    \centering
    \includegraphics[width=\columnwidth]{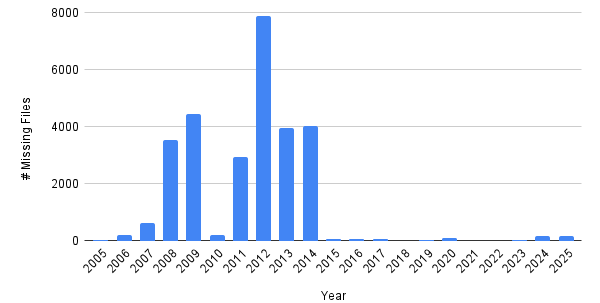}
    \caption{Number of missing files per year from 2005 to 2025}
    \label{fig:missing}
\end{figure}

The comprehensive processing of the ORION Telescope archive data was able to identify and document data outages for the period of  October 2005- June 2025. The analysis found a total of 28,549 missing pcap.gz files, and 1,579 corrupted files from October 2005 to June 2025. Of the 28,549 missing files, 28\% of them were in 2012 while over half are in the range from 2011 through 2014, as shown in Fig.~\ref{fig:missing}. Future studies could investigate correlations between these outages and telescope infrastructure issues, such as connectivity or hardware failures.


\subsection{Data Processing Constraints in Coarse-to-Fine Sub-Pipelines}

We encountered several challenges with processing such a large dataset which consists of massive PCAP files, summarized as follows:
\begin{itemize}
    \item \textbf{Efficient file access}: the compressed pcap.gz files occupy over 390TB of disk space and are stored on an archive server accessed over NFS. Due to the overhead and delays of NFS processing scripts were observed to be spending as much as half of their time waiting for the file to be accessed. To mitigate this issue, files were copied to the local processing server beforehand. However the file-system still bottle-necked the pipeline when trying to access multiple files in parallel. Running the pipelines on files sequentially proved faster than trying to process multiple files in parallel.

    \item \textbf{Handling zipped files} Further compounding the processing time is the time to decompress the file. Native implementations of the gzip utility are single-threaded and cannot handle decompressing large files in a reasonable amount of time. \texttt{rapidgzip} can decompress files in a fraction of the time it takes gzip - up to 75x faster. Apache Drill provides fast and efficient access to PCAP files but does come with a few limitations: unlike the vast majority of other PCAP analysis tools, it cannot process gziped pcaps; and while it does provide access to the packet data it is in an encoded form and does not have the capabilities to extract layer 7 application details from the packet data.

    \item \textbf{Robust import into databases}: importing the processed data into InfluxDB or MariaDB also proved to be a challenge. InfluxDB uses a HTTP API that copies the entire request into memory before writing the data into the database - thus requiring all it and all downstream servers to be configured to accept larger payloads. MariaDB surprisingly took 10x longer to import 20MB csv Files than to import 5MB files. 

\end{itemize}
\subsection{Impact of IP Space Reduction}
In 2018, the Merit network telescope reduced its monitored IP address space from a \texttt{/8} to a \texttt{/13}, consisting of around 500,000 dark IPs, representing a ~60\% decrease in address space. This change had several measurable effects:

\begin{itemize}
\item The number of unsolicited packets captured per day dropped temporarily, likely influenced by the smaller IP exposure.
\item Some scanning tools or botnets may have focused on specific subranges within the \texttt{/8}, so the reduced \texttt{/13} range may no longer intersect these hotspots.
\item Long-term graphs of traffic volume, protocol usage, or scanning rates could be misleading, as changes may reflect telescope scaling rather than global behavior. Normalizing traffic for long-term studies would require further validation.
\end{itemize}

Where applicable, we distinguish between the pre-reduction (\texttt{/8}) and post-reduction (\texttt{/13}) periods when analyzing trends, spikes, or anomalies.

\section{Conclusion and Future Work}
\label{sectionV}
This work presented an initial longitudinal analysis of unsolicited Internet traffic collected over the period (2005–2025) by one of the largest and most persistent network telescopes in the United States, operated by Merit Network. The dataset provides a unique view into global threat activity as observed through scanning and backscatter traffic—key indicators of probing behavior, service enumeration, and ongoing denial-of-service campaigns. In this work, we adopt a coarse-to-fine methodology where insights are first extracted through a more resources efficient pipeline that considers coarse attributes, followed by a second pipeline that extracts finer details but is more
resource extensive. The methodology establishes two pipelines 1) high-level metadata pipeline, and 2) packet header pipeline to process ~20 years of telescope data and extract initial insights. Initial insights show general trends across the period 2005-2025 and focused initial analysis on the year 2024. This study with the established pipelines provides the foundations for further analysis following the proposed coarse-to-fine methodology leveraging existing telescope data (2005-2025) in an optimized fashion.

\section*{Acknowledgment}
Authors would like to acknowledge the contributions of the following individuals for their valuable input during work discussion: Omar ElRefai and Calvin Hanks with Texas A\&M University. 
\bibliographystyle{IEEEtran}
\bibliography{bib}

\end{document}